\documentstyle[referee]{mn}

\newif\ifAMStwofonts
\AMStwofontstrue

\ifoldfss
  \newcommand{\rmn}[1] {{\rm #1}}

  \ifCUPmtlplainloaded \else
    \NewTextAlphabet{textbfit} {cmbxti10} {}
    \NewTextAlphabet{textbfss} {cmssbx10} {}
    \NewMathAlphabet{mathbfit} {cmbxti10} {} 
    \NewMathAlphabet{mathbfss} {cmssbx10} {} 
  \fi
  \ifAMStwofonts
    \ifCUPmtlplainloaded \else
      \NewSymbolFont{upmath} {eurm10}
      \NewSymbolFont{AMSa} {msam10}
      \NewMathSymbol{\upi}     {0}{upmath}{19}
      \NewMathSymbol{\umu}     {0}{upmath}{16}
      \NewMathSymbol{\upartial}{0}{upmath}{40}
      \NewMathSymbol{\leqslant}{3}{AMSa}{36}
      \NewMathSymbol{\geqslant}{3}{AMSa}{3E}

      \let\leq=\leqslant 
       
    \fi
  \fi
\fi 

\ifnfssone
  \newmathalphabet{\mathit}
  \addtoversion{normal}{\mathit}{cmr}{m}{it}
  \addtoversion{bold}{\mathit}{cmr}{bx}{it}
  \newcommand{\rmn}[1] {\mathrm{#1}}

  \newmathalphabet{\mathbfit} 
  \addtoversion{normal}{\mathbfit}{cmr}{bx}{it}
  \addtoversion{bold}{\mathbfit}{cmr}{bx}{it}
  \newmathalphabet{\mathbfss} 
  \addtoversion{normal}{\mathbfss}{cmss}{bx}{n}
  \addtoversion{bold}{\mathbfss}{cmss}{bx}{n}
  \ifAMStwofonts
    \ifCUPmtlplainloaded \else
      \UseAMStwoboldmath
      \makeatletter
      \new@mathgroup\upmath@group
      \define@mathgroup\mv@normal\upmath@group{eur}{m}{n}
      \define@mathgroup\mv@bold\upmath@group{eur}{b}{n}
      \edef\UPM{\hexnumber\upmath@group}
      \new@mathgroup\amsa@group
      \define@mathgroup\mv@normal\amsa@group{msa}{m}{n}
      \define@mathgroup\mv@bold\amsa@group{msa}{m}{n}
      \edef\AMSa{\hexnumber\amsa@group}
      \makeatother
      \mathchardef\upi="0\UPM19
      \mathchardef\umu="0\UPM16
      \mathchardef\upartial="0\UPM40
      \mathchardef\leqslant="3\AMSa36
      \mathchardef\geqslant="3\AMSa3E

      \let\leq=\leqslant 

    \fi
  \fi
\fi 

\ifnfsstwo
  \newcommand{\rmn}[1] {\mathrm{#1}}

  \DeclareMathAlphabet{\mathbfit}{OT1}{cmr}{bx}{it}
  \SetMathAlphabet\mathbfit{bold}{OT1}{cmr}{bx}{it}
  \DeclareMathAlphabet{\mathbfss}{OT1}{cmss}{bx}{n}
  \SetMathAlphabet\mathbfss{bold}{OT1}{cmss}{bx}{n}
  \ifAMStwofonts
    \ifCUPmtlplainloaded \else
      \DeclareSymbolFont{UPM}{U}{eur}{m}{n}
      \SetSymbolFont{UPM}{bold}{U}{eur}{b}{n}
      \DeclareSymbolFont{AMSa}{U}{msa}{m}{n}
      \DeclareMathSymbol{\upi}{0}{UPM}{"19}
      \DeclareMathSymbol{\umu}{0}{UPM}{"16}
      \DeclareMathSymbol{\upartial}{0}{UPM}{"40}
      \DeclareMathSymbol{\leqslant}{3}{AMSa}{"36}
      \DeclareMathSymbol{\geqslant}{3}{AMSa}{"3E}

      \let\leq=\leqslant 

    \fi
  \fi
\fi 

\ifCUPmtlplainloaded \else
  \ifAMStwofonts \else 
    \def\upi{\pi}
    \def\umu{\mu}
    \def\upartial{\partial}
  \fi
\fi

\def\tef{{\it T}$_{\rmn eff}$ }
\def\pop1{Pop.~{\small I}}

 \title[ $UBV$ stellar photometry in M5. I. Peculiarities in the HB stellar distribution]
    {$UBV$ stellar photometry of bright stars in GC M5. I. $UV$ colour-magnitude and colour-colour diagrams and some peculiarities in the HB stellar distribution}

\author[H. Markov et al.]
{H.S.~Markov,$^1$\thanks{Send offprint requests to: H. Markov} N.M.~Spassova,$^2$ and P.V.~Baev$^2$\\
$^1$Inst. of Astronomy, Bul. Acad. of Sci., Natl. Astr. Obs. Rozhen, p.o. box 136, BG-4700 Smolyan, Bulgaria \\
e-mail: rozhen@mbox.digsys.bg\\
$^2$Inst. of Astronomy, Bul. Acad. of Sci., 72 Tsarigradsko ch., BG-1784, Sofia, Bulgaria\\ 
e-mail: neda@astro.bas.bg, baevs@ms3.tu-varna.acad.bg}

\date{Accepted........,  Received 2000 June 1; in orig. form 1999 June 30 }

\pagerange{\pageref{firstpage}--\pageref{lastpage}}
\pubyear{1994}

\begin{document}

\maketitle

\label{firstpage}

%
%

\begin{abstract}
We present stellar photometry in the $UBV$ passbands for the globular cluster  M5$\equiv$NGC5904. The observations, short-exposured photographic plates and CCD frames, were obtained in the RC-focus of the 2m telescope of the Natl. Astron. Obs. 'Rozhen'. All  stars in an annulus with radius  $1\leq r\leq 5.5$ arcmin were measured. We show that the $UV$ CMDs describe different evolutionary stages in a better manner than the 'classical'  $V,\,B-V$ diagram. We use HB stars, with known spectroscopic \tef\,  to check the validity of the colour zero-point. A review of all known $UV$-bright star candidates in M5 is made and some of their parameters are catalogued. Six new stars of this kind are suspected on the basis of their position on the CMD. New assessment of the cluster reddening and metallicity is done using the $U-B,\,B-V$ diagram. We find  [Fe/H]=--1.38, which confirms the Zinn \& West (1984) value contrasting with recent spectroscopic estimates.  In an effort to clarify the question of the gap in the BHB stellar distribution and to investigate some other peculiarities, we use the relatively long-base colour index $U-V$. A comparison of the observed $V,\, (U-V)_0$ distribution of HB stars with a canonical ZAHB model (Dorman et al. 1993) reveals that the hottest stars rise above the model line. This is similar to the '$u$--jump' found in the Str$\ddot{\rmn o}$mgren photometry (Grundahl et al. 1998, 1999). 18 BHB stars with ($B-V)_{0}\,\in [-0.02 \div 0.18$] are used to estimate  their ultraviolet deficiency. It is shown that low gravity ($\log g\leq2$) Kurucz's atmospheric models fit well the observed distribution of these stars along the two-colour diagram.
\end{abstract}
\section{Introduction}

The photometric works on Globular Clusters (GCs), very often,  do not include the $U$ band observations. The $U$--band, however, is very important since it provides a comparatively long--base colour index $U-V$,  which removes the degeneracy in HB colours for $(B-V)_{0}<0$ and is also the most sensitive temperature parameter for early type stars in the $UBV$ system (Buser \& Kurucz 1978). This colour index is furthermore useful to obtain stellar atmospheric parameters using theoretical models (e.g. Kurucz 1992). Taking in mind the good resolution in the RC--focus (F/8, scale 12.86 arcsec\,mm$^{-1}$)  of the 2-m telescope of the Natl. Astron. Obs. 'Rozhen', some years ago, we started a program to study the central parts of some globular clusters. Here are presented the results concerning M5$\equiv$NGC5904. In this case, our aim was to produce $UBV$ magnitudes for a large, statistically significant sample of stars at the level of  the horizontal branch and above and to investigate some puzzling CMD features. We draw special attention to stars occupying the blue horizontal branch (BHB). The distribution of these stars poses  a number of questions that are still open and need further investigation as e.g.:
the presence of a gap in the BHB; the ultraviolet deficiency of some BHB stars; the real location of the bluest HB stars on the $\log L, \log T_{\rmn eff}$ diagram. 

In the following we present the observational material and reduction procedures (Sect.2). The $UV$--diagrams are analysed in Sect. 3. The main topics in this section are the presence of a gap, the $UV$-bright star candidates and the discrepancy between the observed BHB and the zero-age horizontal branch (ZAHB) model. The two--colour diagram features are discussed in Sect. 4. This includes the observed ultraviolet deficiency of BHB stars and the ultraviolet excess of the RGB stars. An assessment of the cluster metallicity and reddening is also given. The results obtained are summarised and discussed in Sect. 5 . 
\section{observational material}

The present investigation of M5 is based on photographic plates and CCD frames taken of the central part of the cluster. To achieve better resolution of stellar images all observations were obtained with short-time exposures. The average seeing, measured at FWHM of the stellar image, is $\sim 2$ arcsec. On each plate - 5$U$,\,6$B$ and  5$V$ - nine overlaping regions, covering up to 5.5 arcmin away from the cluster center, were scanned with a Joyce--Loebl microdensitometer. The data were smoothed by means of a wavelet transform and subsequently converted into relative intensities using calibration curves specific for each plate (for details see Markov 1994). The digital images were analysed with originally developed reduction procedures. To check the reliability of these procedures part of the observations were processed via DAOPHOT (Stetson 1987) -- a detailed comparison of the results is given in Markov et al. (1997). The efficiency of the wavelet transform in localising stellar images in crowded fields is described in Borissova et al. 1997. A similar approach had been previously employed by Auri\'ere \& Coupinot (1989) to obtain  improved photometry for stars in dense globular clusters fields. 
%
%
In order to obtain more precise photometric calibration, supplementary 3$V$ and 3$B$ CCD observations of the M5 core were taken at March 16, 1997 with a SBIG ST-6 camera (pixelsize = 0.32 arcsec). Seeing conditions for the night were stable with FWHM$\sim 0.9$. The standard field in M92 of  Christian et al. (1985) was observed to calibrate the program images.

The transformation of the instrumental magnitudes to the standard $U$ system was performed by means of 24 stars from the Arp's (1962) photoelectric standard sequence in the field of M5. This sample covers wide range of colours:
$-0.23<B-V<1.48$, 
$-0.32<U-B<1.56$ and 
$-0.55<U-V<3.04$.
The BHB stars are of special interest in our investigation and separate transformations to the standard system have been employed for them. In this colour region 7 Arp's photoelectric and 3 secondary photometric standard stars, taken from Richer and Fahlman (1987), were used.

The calibration into $B,\,V$ standard system was performed in two steps. First, we used the CCD observations to calibrate $BV$ instrumental magnitudes with respect to the primary standard stars in M92 (Christian et al. 1985). Then, 86 stars common for photographic and CCD measurements were used as a local secondary standard to calibrate all other program stars. The mean internal errors are in the range of $0.02\div 0.04$. The total error including the uncertainties in the photoelectric data is $\pm0.03$ in the stellar magnitudes and $\pm0.04$ in the colour indices. 

Our photometry can be compared with earlier works e.g.:
\begin{enumerate}
\item  {\it the UBV CCD photometry of von Braun et al. (1998)}. For 18 stars in common, the mean differences between our data and those of von Braun et al. (1998) are: 
$\langle\Delta U\rangle = 0.019\pm 0.032$; 
$\langle\Delta(U-B)\rangle = 0.013\pm 0.031$ and 
$\langle\Delta(U-V)\rangle = 0.036 \pm 0.038$. 
Obviously, the two magnitude sets match each other within the error. A systematic difference of about 0.06 mag was found for the U magnitudes of 5 AGB stars (in the sense von Braun's data being brighter). These stars were not included in the above assessments. 
\item {\it the photographic photometry of BCF}. The mean difference in the zero point ($V_{\rm our} - V_{\rm BCF}$), based on 182 common stars, is 0.01 mag. There seems to be a small colour difference of 0.03 mag for BHB stars  (our magnitudes being redder).
\item {\it the CMD mean lines of Sandquist et al. (1996)}. Figure 1 shows the $V, B-V$ diagram on which the Sandquist et al. (1996) fiducial lines are overlaid for comparison. 
\end{enumerate}
\section{$UV$ colour--magnitude diagrams}
\subsection{The $V, U-V$ diagram} 

Figure 2a shows the $V, U-V$  diagram for M5. Two features of this diagram are worth noting: the BHB in this plane is not as steep as that in $V, B-V$ (Fig. 1); the BHB stars form an almost linear sequence with a discontinuity at $V\approx 15.45$ and $U-V\approx 0.05$. The latter is illustrated in the histogram (Fig. 2a). To test statistical significance of this BHB feature, the $\ell_{\rmn HB}$ co-ordinate deduced from Ferraro et al. (1992) and Dixon et al. (1996) is used. $\ell_{\rmn HB}$ is a measure of the star position along the HB in the $M_{\rmn V}, (U-V)_0$ diagram as marked by its projection on the HB ridgeline. The $\ell_{\rmn HB}$ parameter grows from the red extreme (RE) to the blue end of the HB. The zero point for $\ell_{\rmn HB}$ is set at $(U-V) _{0}^{\rmn RE}=0.7$. The HB ridgeline is determined as a non-linear fit to the sample of 160 HB stars supplemented with 68 RR Lyr variables. Their colours are assumed to be randomly distributed inside the instability strip boundaries $(0.26>(U-V)_{0}<0.45)$ with a mean magnitude $\langle V_{\rm RR}\rangle = 15.19$. The scale ratios used is  $\Delta \ell_{\rmn HB}/ \Delta(U-V)_0=3.341$ and $\Delta \ell_{\rmn HB}/ \Delta M_{\rmn V}\approx 1.42$.  A histogram of the $\ell_{\rmn HB}$ distribution is shown in Fig. 3 and the cumulative distribution -- in Fig. 4. The gap is clearly recognised at $\ell_{\rmn HB}=0.42\pm0.057$. Following Hawarden (1971) and Newell (1973), we tested the significance of the gap by means of '$\chi^2$ statistics' applied to the cumulative distribution and found that the gap is statistically significant at the 99.93 per cent level. However, Catelan et al. (1998) have recently argued that the Hawarden-Newell technique substantially overestimates the probability that the gap is real. This result together with the fact that only 80 per cent (228 stars) of all observed HB stars have been involved in our analysis make the reality of the gap not convincing enough.

Figure 5 shows the theoretical ZAHB (Dorman, Rood, O'Conel 1993, oxygen-enhanced model, $Y_{\rmn HB}=0.249$; [Fe/H]=-1.48; [O/Fe]=0.63; $(m-M)_{\rmn V,0}=14.46$) overlaid on the observed HB. Squares denote stars with spectroscopically derived \tef (Crocker, Rood \& O'Connel 1988). The compatibility of both sets is satisfactory. Fig 5. shows that although practically all HB stars lie above the theoretical sequence the discrepancy become marked above
 $T_{\rmn eff}\approx 10\,800\pm700\,{\rmn K} ((U-V)_{0}\approx-0.025)$. Grundahl, Catelan, Landsman, Stetson \& Andersen (1999) observed a similar effect (called '$u$--jump') in 14 GC (including M5) from $u,\,y$  photometry. These authors suggested that '...the $u$--jump is a ubiquitous feature, intrinsic to {\it all\,} HB stars hotter than $T_{\rm tef}\approx 11\,500\, {\rm K}$'. 

\subsection{The $U,\,U-V$ diagram } 

We note the following points regarding the  $U,\,U-V$ diagram (Fig. 2b):
\begin{enumerate}
\item The stars of the upper part of the RGB are as bright as the BHB stars. The $U$ magnitudes of the brightest RGB stars increase as the stars become redder.
\item The AGB is clearly distinguished from the RGB being, in average, more than 0.20 mag brighter. Note that the brightest AGB stars have an astrometric membership probability $P = 99$ per cent (Rees 1993; Spassova \& Michnevski 1981). The star on the AGB tip, denoted with full square, is ZNG2 $UV$--star (Zinn et al. 1972).
     \begin{table}
     \caption{Data for {\it UV}--bright stars candidates detected in GC M5}
     \begin{center}
     \begin{tabular}{ccccccc}
     \hline
     \hline
     $\#$  &identification    &U   &V &B-V &status  &ref.\\
     \hline
     \\
     1&   ZNG1&             --&    --&      --&  --&  --\\
     2&   ZNG2&           14.90&  12.01&  1.55&   m&   1\\
     3&   ZNG3, III-24&   .....&  13.51&  0.69&   f&   1\\
     4&   ZNG4,  IV-78&   14.44&  13.62&  0.78&   f&   1\\
     5&   ZNG5,  IV-65&   14.49&  14.05&  0.51&   f&   1\\
     6&   ZNG6, var-41&    --&     --&      --&   m&   1\\
     7&   ZNG7, var 25&    --&     --&      --&   m&   1\\
     8&   III-37&         15.04&  14.79&  0.02&   m& 2,3\\
     9&   B169&           15.09&  14.61&  0.38&   m&   3\\
    10&   our 62&         15.09&  15.27& -0.02&  --&   4\\
    11&   I-17&           15.18&  14.96&  0.04&  --&   4\\
    12&   I-44&           15.21&  15.07&  0.07&   m&   4\\
    13&   II-22&          15.18&  15.23&  0.01&  --&   4\\
    14&     --&            --&    14.56&  0.42&  --&   4\\
    15&     --&            --&    14.18&  0.60&  --&   4\\
     \hline
     \\
     \end{tabular}
     \begin{list}{}{}
     1. Zinn et al. (1972); 2. Cudworth (1979); 3. Rees (1993); 4. this study
     \end{list}
     \end{center}
     \end{table}
\item The group of the so called '$UV$--bright stars' is clearly separated from the principle sequences. They are indicated with full squares in Fig. 2. Six of the seven $UV$--bright stars from Zinn et al. (1972) are measured in our study. Three of them are field stars, and one is variable 41 (Sawyer Hogg 1973). The star ZNG7 is probably variable 25. On the frames it appears blended with the RGB star III--140 (Buonanno et al. 1981). Our $U$--photometry confirms that the cluster member III--37 (Cudworth 1979) is an $UV$--bright star: although both magnitudes and colours show big dispersion in all studies its position on the classical 
$V, B-V$ diagram is above the blue HB. Rees (1993) noted as other possible $UV$--bright stars: III--48 (Arp 1962), B169 and B200 (Barnard 1931). Our results confirm this status only for B169. We suspected four new possibles $UV$--bright stars: I--17, I--44, II--22 and the star numbered 62 in our catalogue. Another two stars were added to the list according to their position on the {\it V,\,B--V\,} diagram. These two stars, with magnitudes and colours: $V=14.56$, $B-V=0.42$ and $V=14.18$, $B-V=0.60$,  were detected in the central part of the cluster. Thus, 15 $UV$--bright stars candidates are found by different authors in M5. Table 1 lists $UV$--bright stars including some that probably belong to the so-called 'blue nose' ($f\equiv$\,field, $m\equiv$\,cluster member).
\end{enumerate}
\section{ The colour--colour diagram}

The $U-B,\,B-V$ relation can be used to determine the foreground reddening towards M5. Our intrinsic colours were derived assuming reddening E$_{\rmn B-V}$=0.03 mag, with a slope E$_{\rmn U-B}$/ E$_{\rmn B-V}$=0.72. 
For 38 BHB stars with {\it B--V}$<$--0.02 we estimate E$_{\rmn B-V}$=0.03$\pm$0.014.  The spectroscopically determined \tef and $\log g$ for 11 HB stars in M5 (Crocker et al. 1988), also allow us to determine the cluster reddening.  The $(B-V)_0$ values for these stars were derived using corresponding relations found in Kurucz (1992). In this way we obtained ${\rmn E}_{\rmn B-V}=0.029\pm0.015$.

Figure 6 shows the two--colour diagram for RGB (left panel, crosses) and AGB (right panel, triangles) for stars with {\it V} magnitudes above the level of the HB ($V\gid 15.09$). In the both panels, the unreddened intrinsic two--colour line for \pop1 normal (L.C. {\small III}) giants (Buser \& Kurucz 1992) is drawn (solid line). The two types of stars show the well--known ultraviolet excess $\langle\Delta$({\it U--B})$_{0}\rangle$. This excess, measured at  ({\it B--V})$_{0}$=1.0, is a sensitive indicator for the metal abundance (Wallerstein \& Helfer 1966; Sandage 1970). In our case $\Delta$({\it U--B})$_0$ for all stars with ({\it B--V})$_{0} \in$\,[0.92$\div$1.08] were averaged. The mean excess $\Delta$({\it U--B})$_{1.0}$ is 0.178$\pm$0.02. Using the Buser \& Kurucz's relation between $\Delta$({\it U--B})$_{1.0}$ and [M/H] we obtained 
[Fe/H]$=-1.38\pm0.09$. The quoted error  is derived only from the formal error propagation in the photometric and model data. The M5 metallicity assessments found in the literature show a wide range (Briley et al. 1992, their Table 3). Our value is in an agreement with that found in Zinn \& West (1984), $-1.4$, but is inconsistent with the more recent high-resolution spectroscopic measurements carried out by Carreta \& Bragaglia (1998), $-1.11\pm0.11$; Carreta \& Gratton (1997), $-0.9$, and Sneden, Kraft, Prosser and Langer (1992) $-1.17\pm0.01$. While we could not offer a certain answer on this topic, we would like to note that Borissova et al. (1999) pointed out a remarkable similarity of the upper red giant branches of M5 and NGC6229. For this reason only, one might expect both clusters to share the same metallicity. The reliability of the value derived photometrically for NGC6229 in Borissova et al. (1999), [Fe/H]$=-1.4$, (based on the Zinn \& West scale) is confirmed by Wachter, Wallerstein, Brown and Oke (1998). These authors obtained the global metallicity of the cluster,[M/H] $=-1.4$, from a direct medium-resolution spectroscopy of cluster giants.

The AGB stars (within the same colour range) show a smaller ultraviolet excess in comparison with the RGB stars. This is demonstrated in Fig. 6 (right panel), where the mean position of the RGB stars is marked with a dashed line. The difference in the $UV$--excesses for RGB and AGB stars is obvious. The implied difference in $\Delta(U-B)_0$ is 0.10 mag at $(B-V)_0=1.0$. A smaller $UV$ excess of  the AGB stars has been found practically in all globular clusters with available $UBV$ measurements. 

Figure 7 illustrates the HB in the colour--colour plane based on 161 stars (114 BHB and 47 RHB). A dashed line shows the empirical \pop1 line. A solid line indicates the model ZAHB transformed to $(B-V)_{0},\,(U-B)_{0}$. This two--colour diagram reveals the following features:
\begin{itemize}
\item Stars with $(B-V)_{0}<-0.02$  fit very well the empirical \pop1 relation and ZAHB. These are stars occupying the region in the $(B-V)_{0},(U-B)_{0}$ plane, where differences in {\it g} and metal abundance have negligible effect on the observed colours. 
\item The cool BHB stars, appearing just to the blue side of the RR Lyrae instability strip, show a small but definite ultraviolet deficiency with respect to the intrinsic two--colour line (Allen 1973). The mean deficiency measured for 18 stars in the colour interval $-0.01\leq(B-V)_{0}\leq0.18$ is $\Delta(U-B)_{0}=0.102\pm0.019$. This effect was first found by Arp (1962)  and is based on only 5 BHB stars with photoelectric measurements. The same effect was also observed in some other clusters: NGC 1904 (Stetson \& Harris 1977; Kravtsov et al. 1997), NGC 3201 (Lee 1977), NGC 6723 (Menzies 1974), NGC 5139 (Newell et al. 1969b), NGC 6397 (Newell et al. 1969a), M 15 (Sandage 1970), NGC 1841 (Alcaino et al. 1996). To quantify the effect, we reanalysed available {\it UBV} data for BHB stars in some of these clusters using the E$_{\rmn B-V}$ values found in Harris (1997). Table 2 summarises the results.
     \begin{table}
     \caption{ The ultraviolet deficiency for cool BHB stars detected in some globulars }
     \begin{center}
     \begin{tabular}{lllll}
     \hline
     \hline
     \\
     NGC  &$\delta$ (U-B)    &n   &E$_{\rmn B-V}$ &reference
     \\
     \hline
     \\
        1904 &0.162 $\pm 0.028$ &20 &0.01 &Stetson et.al., 1977\\
        3201 &$0.136 \pm 0.014$ &26 &0.21 &Lee, 1977\\
        5139 &$0.087 \pm 0.035$ &6  &0.12 &Newell et.al., 1969b\\
        5904 &$0.102 \pm 0.019$ &18 &0.03 &this study\\
        6397 &$0.085 \pm 0.018$ &8  &0.18 &Newell et. al., 1969a\\
        6723 &$0.162 \pm 0.047$ &12 &0.05 &Menzies, 1974\\
     \\
     \hline
     \end{tabular}
     \end{center}
     \end{table}
The third column shows the number of stars used. However, extensive photoelectric observations of BHB stars in other globular clusters give somewhat conflicting results. Data for M3, M13 and M92 (Sandage 1970) place such stars essentially on the standard \pop1 two--colour line. Although the possibility of a small systematic error in our $U$ magnitudes cannot be totally neglected, the observed effect in the M5 two--colour diagram may be at least partly real. The different values of the ultraviolet deficiency, toghether with it absolute absence in some cases, points to a probable connection of this observable with the cluster parameters. 
%
%

The ultraviolet deficiency of the BHB stars in respect of the canonical ZAHB sequence is smaller (see Fig.~7). Although the scatter of the data is large  -- mainly due to the low quality of the photographic observations -- the deviation from the model line is systematic and appears to be real. From the Kuruzc's (1992) computations it is evident that $(U-B)_0$ colour is a gravity indicator for stars with $T_{\rm tef}<10\,000\,{\rmn K}$ ($(B-V)_{0}>-0.02$). This point is illustrated in Fig.~7, where the $\log~g = 2$ and $\log~g = 4$ model loci are plotted. It is clearly seen, that a satisfactory overlapping of the observed and theoretical sequences is achieved involving low gravity ($\log~g\leq 2.0$) atmospheric models transformed to the 
$(U-B)_{0},\,(B-V)_{0}$ plane.  

The suggestion of Arp (1962) that '... fainter BHB stars, as they get very blue have an increasingly large ultraviolet deficiency relatively to \pop1 line.'  is not confirmed.  
\item The RHB stars do not show any ultraviolet deficiency or ultraviolet excess with respect to the standard \pop1 colour--colour line. It has been known  (Sandage 1969, 1970) that the sequence of metal-deficient RHB stars follows an empirical \pop1 relation. Sandage (1970) gives an explanation in terms of the 'fortuitous' cancellation of the effects of a low metal abundance and low surface gravity.
%
\end{itemize}

\section{ Summary and conclusions}

$UBV$ photographic and $BV$ CCD observations in the central part ($r\leq 5.5$ arcmin) of the globular cluster $M5\equiv NGC5904$ are presented. The $UV$ colour-magnitude and two-colour diagrams for stars down to $V=16.5$ mag are analysed. 

The main results of our analysis can be summarised as follows :
\begin {enumerate}
\item We confirm the presence of a gap in the BHB stellar distribution of M5 announced by Brocato at al. (1995) and Drissen \& Shara (1998).  In the $V,\,U-V$ diagram, the gap is visible at $V\approx15.45$ and $U-V\approx 0.05$. Although the high statistical significance of this BHB feature, we consider the question of its reality still uncertain and will discuss it elsewhere (Baev et 
al., in press).
\item The ultraviolet excess observed for RG stars, 
$\Delta$ ({\it U--B})$_{0, 1.0}$=0.178$\pm$0.02, is used to estimate the metallicity of the cluster. The value obtained in our study, $-1.38\pm0.09$, confirms that derived by Zinn \& West (1984), $-1.4$. 
\item 6 stars are suspected as new  '$UV$--bright' stars. The $UBV$ data for these stars are summarised in Table 1.
\item In the $V,\,(U-V)_0$ diagram we find that the hottest BHB stars tend to lie above the canonical ZAHB at a temperature threshold  $T_{\rm eff}=10\,800\pm 700~{\rm K}$. This resembles the effect observed for 14 globular clusters by Grundahl et al. (1998,\,1999) and called '$u$--jump'. In our investigation, the BHB sample is considerably richer than those used by GCLSA. This result on the other hand demonstrates the ability of Johnson's $UBV$ to reproduce an effect observed by means of the Str$\ddot{\rmn o}$mgren's photometry.
\item Ultraviolet deficiency is detected for cool BHB stars. Based on Kuruzc's (1992) atmospheric models, we interprete this effect as an indication for  lower surface gravity for these stars. Having in mind that  '$u$-jump' demonstrated in GCLSA is intimately connected to low gravities ('$\log g$--jump') a conclusion could be drawn that two groups of BHB stars in M5 tend to show surface gravities lower than those predicted by existing ZAHB models.
\end{enumerate}
Finally, we would like to add few remarks concerning the nature of the ultraviolet deficiency and the jump. GCLSA argued  that a stellar atmosphere effect -- radiative levitation of metals -- rather than helium mixing, is the primary cause of the $u$- and $\log g$-jump phenomenon. One of their arguments against helium mixing is that the size of the jump and its location do not depend on a cluster's individuality. Collecting data for ultraviolet deficiency in clusters with available $UBV$ data, we found that the magnitude of the deficiency varies from cluster to cluster (Table 2) while  in some clusters it is not observed at all. These finds together with the results of GCLSA allow  us to suggest that the jump and the deficiency  more likely have different nature. In addition, taking in view certain evidences for mixing processes in some M5 red giant stars (based on spectral observations referenced in GCLSA), we suppose that the ultraviolet deficiency might be a photometric signature of helium mixing. This draws a possible direction for a future ultraviolet deficiency investigation: to obtain a homogeneous set of UBV data for a large number of GCs and to study the relation of the cluster parameters to the characteristics of the deficiency (if any).
\section*{Acknowledgements}
The authors are grateful to S.Moehler, F.Grudahl and M.Catelan for helpful information and discussions. We would also like to thank the anonymous referee for very careful and detailed review of our paper.\\
This research was supported by the National Sci. Found. grant under contract No.
F-604/1996 with the Ministry of Education and Sciences.\\
This research has made use of NASA's Astrophysics Data System Abstract Service.
\label{Acknowledgements}
\pagebreak

\pagebreak
\begin{center} \textbf{Figure captions}  \end{center}
Fig. 1 $V,\,B-V$ diagram for stars measured in the annulus with radius $1\leq r\leq 5.5$ arcmin. \smallskip\\
Fig. 2. The $UV$ colour-magnitude diagrams. In both diagrams, the place of the suspected gap is indicated by a bar; the $UV$--bright stars are marked with squares. In the panel a), above the BHB, is shown the histogram of the stellar counts found in the corresponding colour bins. 
\bigskip\\
Fig. 3. The $\ell_{\rmn HB}$ distribution histogram.
\bigskip\\
Fig. 4. The $\ell_{\rmn HB}$ cumulative distribution function.
\bigskip\\
Fig. 5. Comparison between the observed BHB (grey points) and the ZAHB (solid line).  Note the general discrepancy between observations and theory as well as the more rapidly rising discrepancy for colours below $((U-V)_{0}\approx-0.025)$. The squares mark the positions of M5 BHB stars with spectroscopically measured \tef and $\log g$ (Crocker et al. 1988). 
\bigskip\\
Fig. 6.  The two colour diagrams for RGB (left panel) and AGB (right panel) stars. The standard Pop. {\small I} line (Allen 1973, solid line) is overlaid on both diagrams. In the right pannel, the mean position of the RGB stars is marked with a dashed line.
\bigskip\\
Fig. 7. Two-colour diagram for HB stars. The empirical \pop1 line is marked with a dashed line.  ZAHB is overlaid also (bold solid line). Two-colour sequences taken from the Kurucz (1992) atmospheric models for two different values of $\log g$ are marked with filled stars and squares.

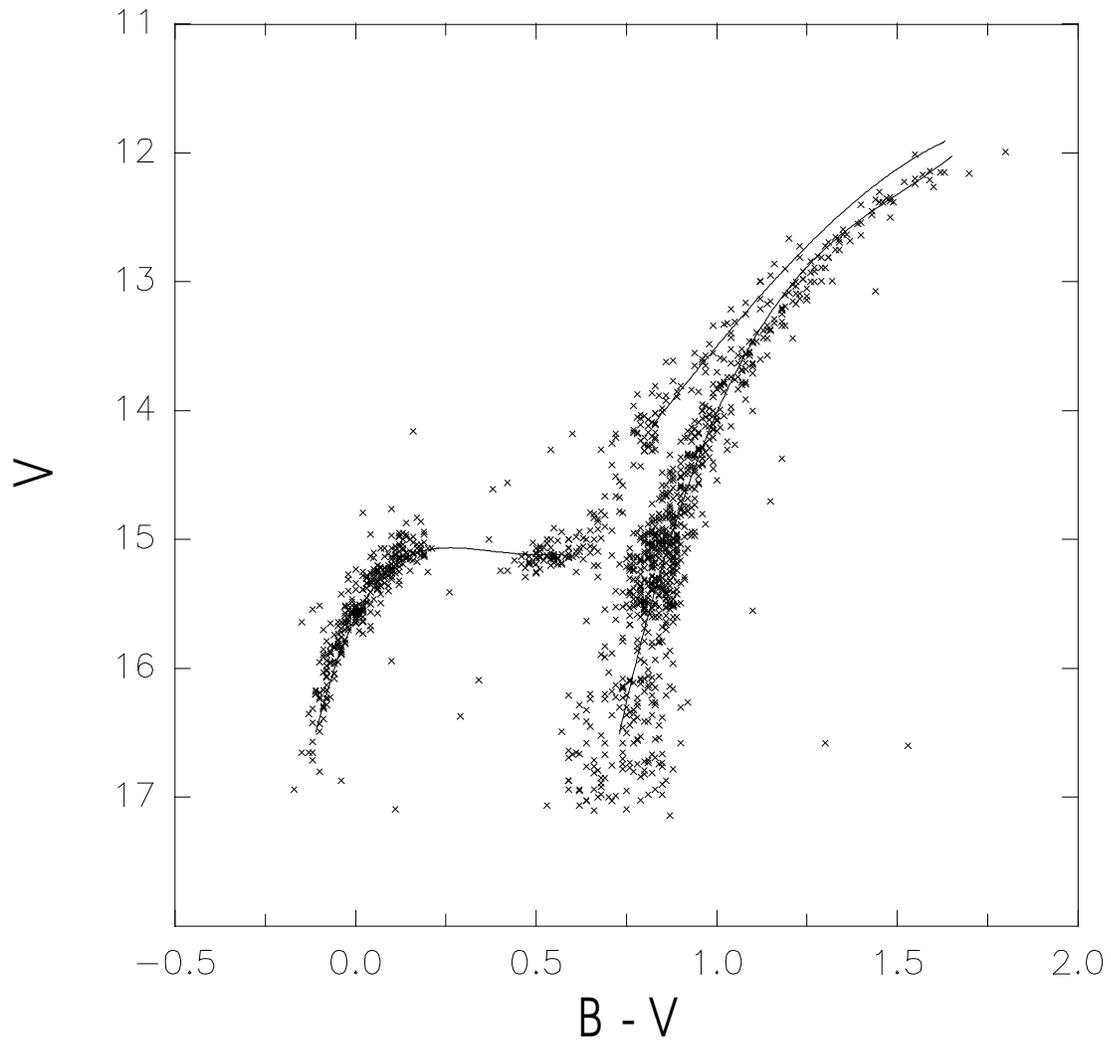
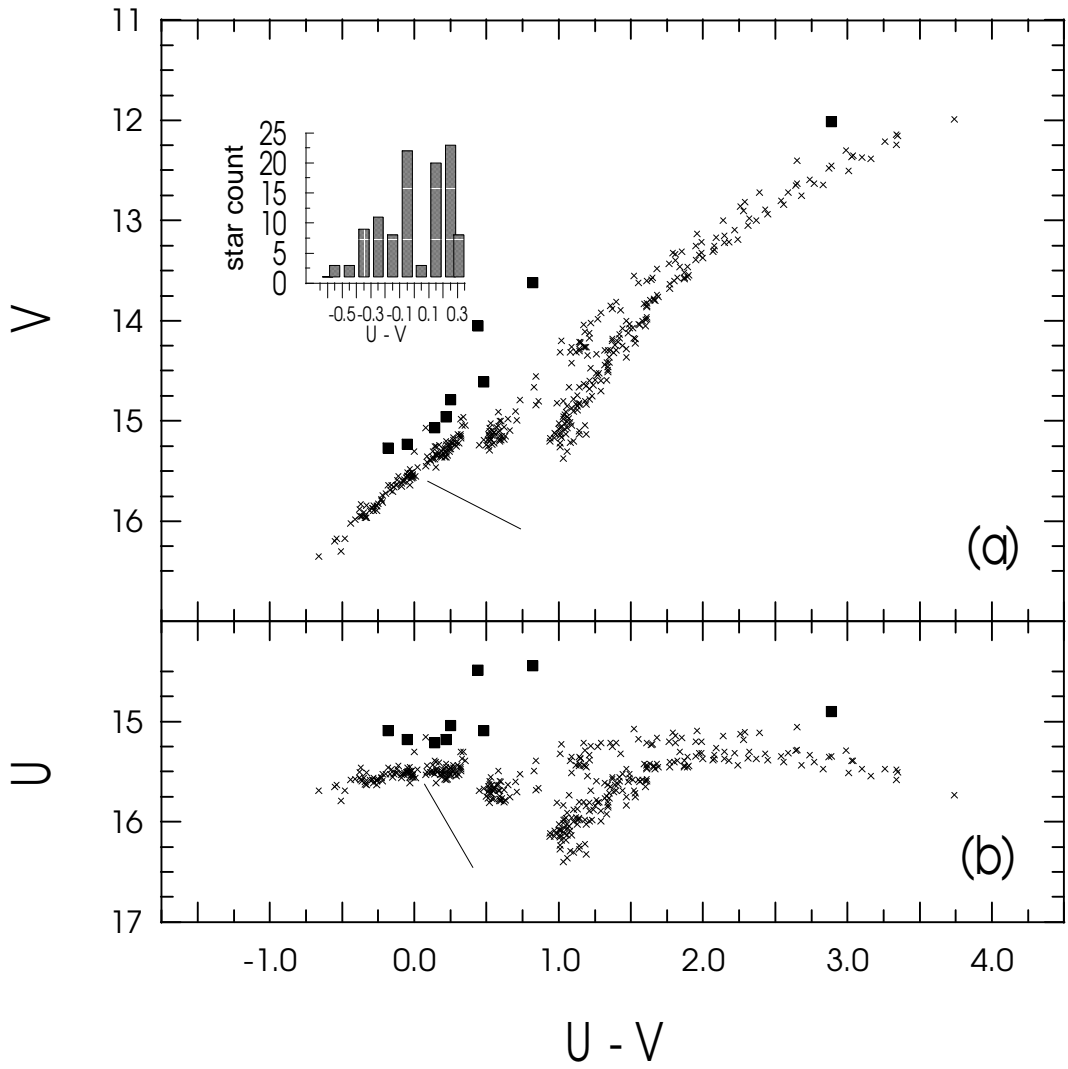
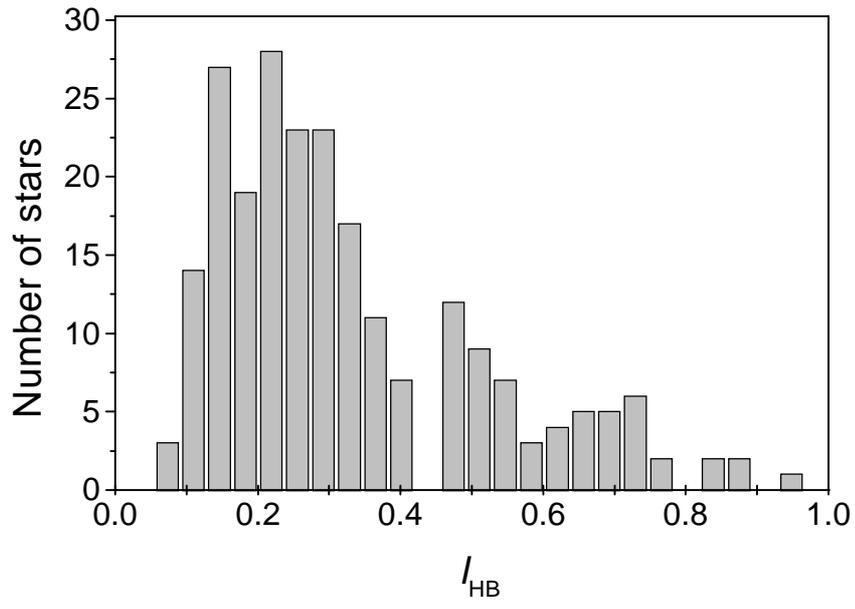
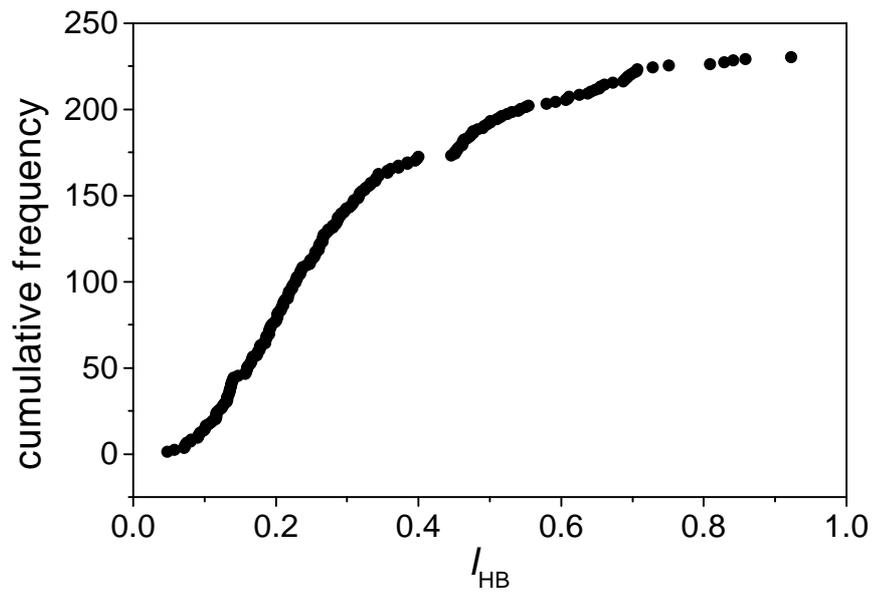
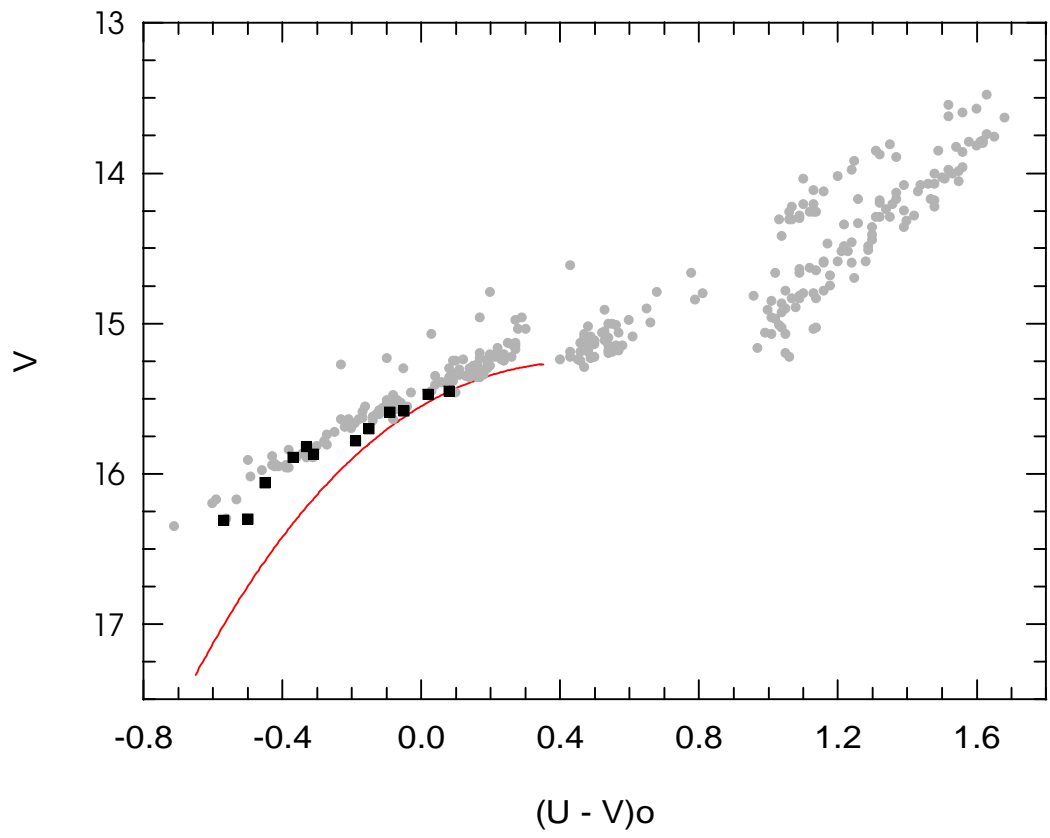
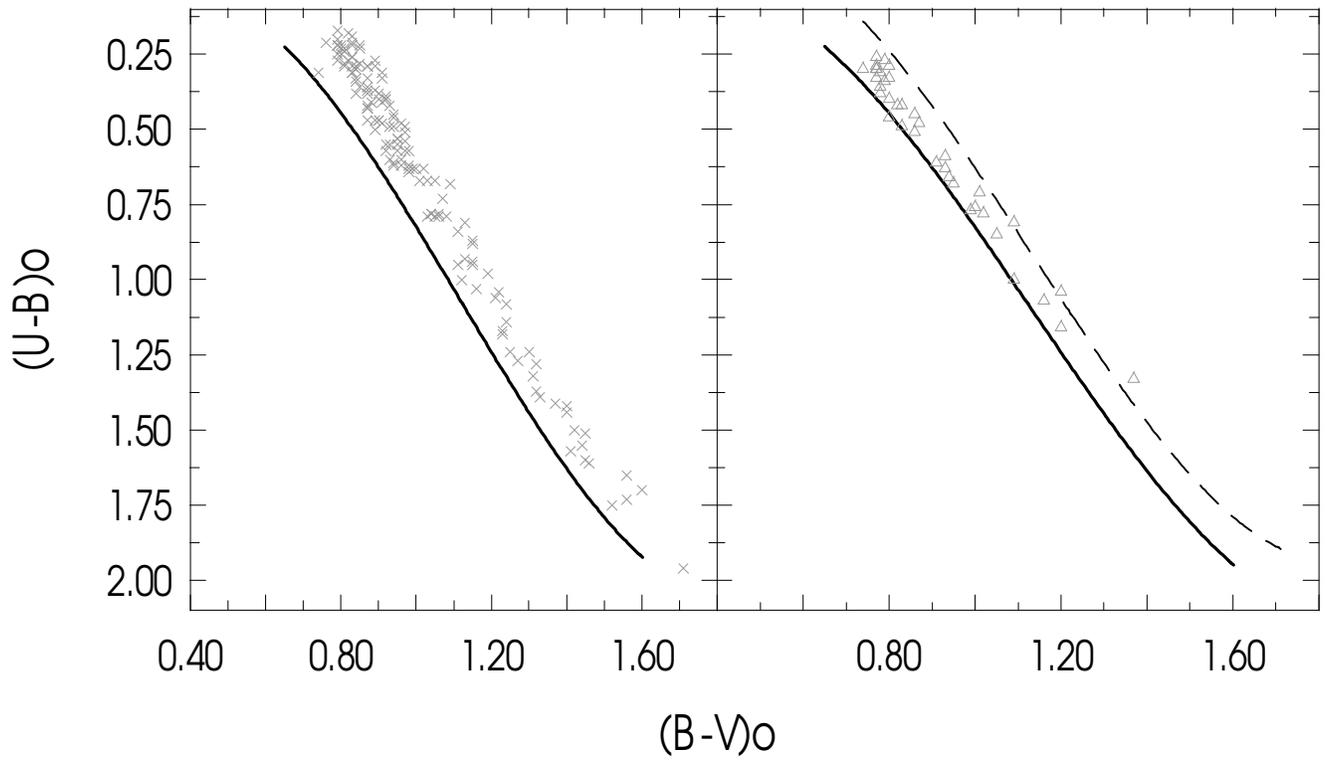
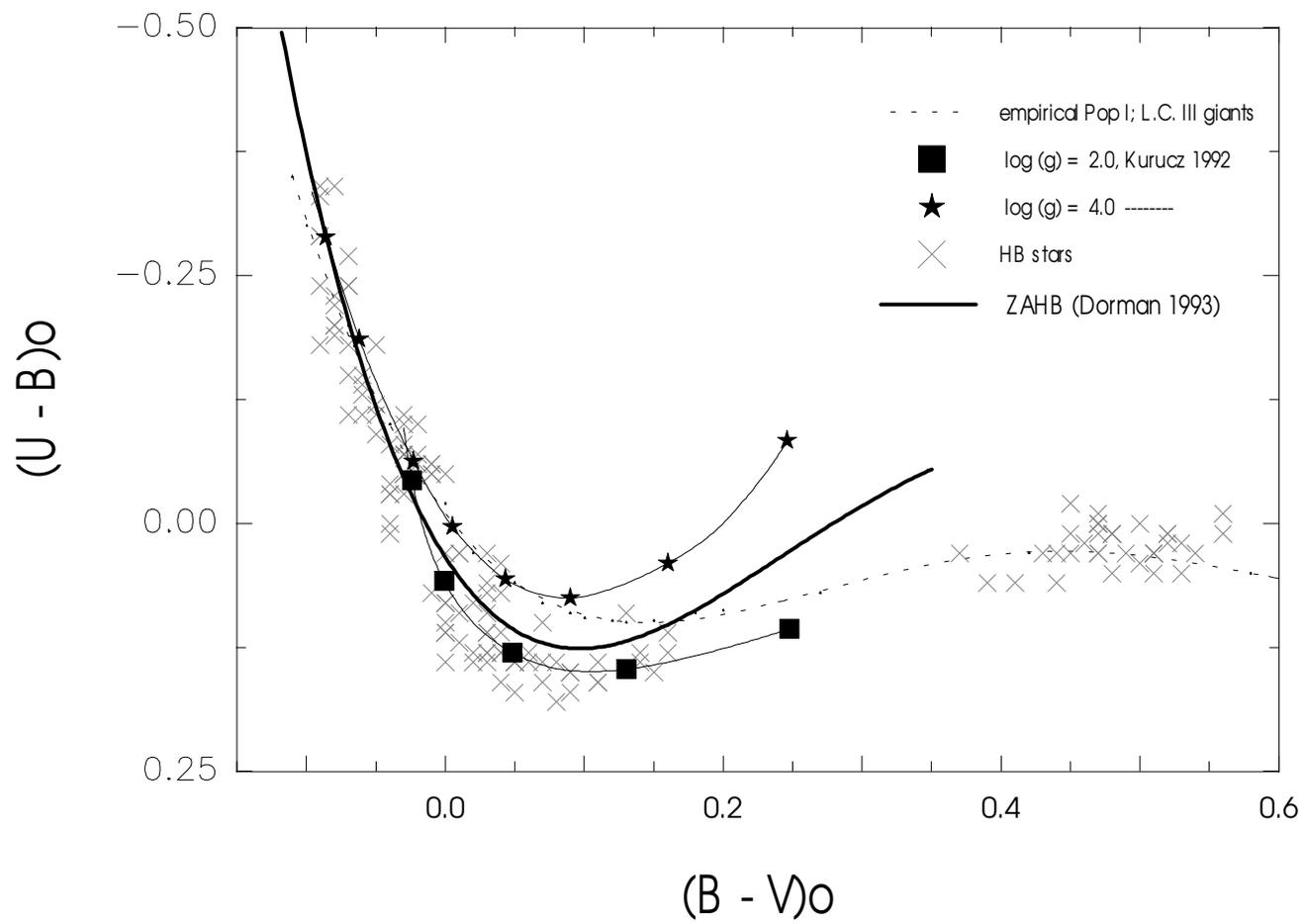
\end{document}